\pdfoutput=1
\documentclass[12pt]{article}

\usepackage{lmodern}

\DeclareFontFamily{T1}{calligra}{}
\DeclareFontShape{T1}{calligra}{m}{n}{<->s*[1.44]callig15}{}
\DeclareMathAlphabet\mathcalligra   {T1}{calligra} {m} {n}
\DeclareMathAlphabet\mathzapf       {T1}{pzc} {mb} {it}
\DeclareMathAlphabet\mathchorus     {T1}{qzc} {m} {n}
\DeclareMathAlphabet\mathrsfso      {U}{rsfso}{m}{n}
\DeclareMathAlphabet\mathfrcal      {T1}{frcursive}{m}{it}
\DeclareFontFamily{T1}{frcursive}{}
\DeclareFontShape{T1}{frcursive}{m}{n}{<->s*[1.44]callig15}{}
\DeclareMathAlphabet\mathfrcal      {T1}{frcursive}{m}{it}

\usepackage{amsmath}
\usepackage{amssymb}
\usepackage{amsthm}
\usepackage{graphicx}
\usepackage[dvipsnames]{xcolor}
\usepackage[english]{babel}
\usepackage{csquotes}
\numberwithin{equation}{section}
\usepackage{array}
\usepackage{mathtools}
\usepackage{dsfont}
\usepackage{mathrsfs}
\usepackage{tikz}\usetikzlibrary{matrix,fit}
\usepackage{varwidth}
\usepackage{enumerate}
 \usepackage{appendix}
\usepackage{ytableau}
\usepackage{simpler-wick}
\usepackage{bm}

\usepackage{caption}

\usepackage[compat=1.1.0]{tikz-feynman}

\usepackage[margin=1in]{geometry}
\usepackage{nicematrix}

\usepackage[
    backend=bibtex,
    style=alphabetic,
maxbibnames=99,
giveninits=true,
minalphanames=1,
maxalphanames=3,url=false
  ]{biblatex}

\usepackage{mathabx}
\usepackage{empheq}

\usepackage{setspace}

\usepackage{fontawesome5}

\usepackage{slashed}
\usepackage{upgreek}

\usepackage{tabstackengine}

\usepackage{wrapfig}
\usepackage[abs]{overpic}

\usepackage{float}

\usepackage{mathtools}

\fixTABwidth{T}

\usepackage{accents}

\newcommand{\bea}{\begin{equation}}
\newcommand{\eea}{\end{equation}}
\newcommand{\bear}{\begin{eqnarray}}
\newcommand{\eear}{\end{eqnarray}}
\newcommand{\bearr}{\begin{eqnarray*}}
\newcommand{\eearr}{\end{eqnarray*}}

\newtheorem{prop}{Proposition}

\newdimen\mytextwidth
\newcommand\rem[2][cyan!40!green]{\noindent\nobreak\hfil\penalty1000\hfilneg
\mytextwidth=\linewidth\advance\mytextwidth by 2mm
\begin{tikzpicture}[baseline=-\the\dimexpr\fontdimen22\textfont2\relax]\node[outer sep=0pt,draw=black,fill=#1,fill opacity=1,text opacity=1,rectangle,rounded corners]{\begin{varwidth}{\mytextwidth}\textcolor{white}{#2}\end{varwidth}};
\end{tikzpicture}\allowbreak
}

\newcommand\whiterem[2][white!]{\noindent\nobreak\hfil\penalty1000\hfilneg
\mytextwidth=\linewidth\advance\mytextwidth by 2mm
\begin{tikzpicture}[baseline=-\the\dimexpr\fontdimen22\textfont2\relax]\node[outer sep=0pt,draw=black,fill=#1,fill opacity=1,text opacity=1,rectangle,rounded corners,line width=1.5pt]{\begin{varwidth}{\mytextwidth}\textcolor{black}{#2}\end{varwidth}};
\end{tikzpicture}\allowbreak
}

\renewbibmacro{in:}{}

\usepackage{mdframed}

\newbibmacro{string+doiurlisbn}[1]{%
  \iffieldundef{doi}{%
    \iffieldundef{url}{%
      \iffieldundef{isbn}{%
        \iffieldundef{issn}{%
          #1%
        }{%
          \href{http://books.google.com/books?vid=ISSN\thefield{issn}}{#1}%
        }%
      }{%
        \href{http://books.google.com/books?vid=ISBN\thefield{isbn}}{#1}%
      }%
    }{%
      \href{\thefield{url}}{#1}%
    }%
  }{%
    \href{http://dx.doi.org/\thefield{doi}}{#1}%
  }%
}

\DeclareFieldFormat{title}{\usebibmacro{string+doiurlisbn}{\mkbibemph{#1}}}
\DeclareFieldFormat[article,incollection]{title}%
    {\usebibmacro{string+doiurlisbn}{\mkbibquote{#1}}}

\newmdenv[
  topline=false,
  bottomline=false,
  rightline=false,
  linewidth=2pt,
  skipabove=\topsep,
  skipbelow=\topsep
]{siderules}

\newmdenv[
  topline=false,
  bottomline=false,
  linewidth=2pt,
  skipabove=\topsep,
  skipbelow=\topsep
]{siderulesright}

\makeatletter
\renewcommand{\@seccntformat}[1]{\csname the#1\endcsname.\quad}
\makeatother

\usepackage{setspace}
\usepackage{xpatch}

\makeatletter
\renewcommand{\@chap@pppage}{
  \clear@ppage
  \thispagestyle{plain}
  \if@twocolumn\onecolumn\@tempswatrue\else\@tempswafalse\fi
  \null\vfil
  \markboth{}{}
  {\centering
   \interlinepenalty \@M
   \normalfont
   \MakeUppercase \appendixpagename\par}
  \if@dotoc@pp
    \addappheadtotoc
  \fi
  \vfil\newpage
  \if@twoside
    \if@openright
      \null
      \thispagestyle{empty}
      \newpage
    \fi
  \fi
  \if@tempswa
    \twocolumn
  \fi
}
\makeatother

\definecolor{navycol}{RGB}{100,150,160}
   \definecolor{pinkcol}{RGB}{242,55,55}
   \definecolor{greencol}{RGB}{50,205,50}

   \definecolor{bluecol}{RGB}{30,144,255}

\usepackage{tocloft}

\usepackage{titlesec}

\titleformat*{\section}{\large\bfseries}
\titleformat*{\subsection}{\normalsize\bfseries}
\titleformat*{\subsubsection}{\normalsize\bfseries}
\titleformat*{\paragraph}{\large\bfseries}
\titleformat*{\subparagraph}{\large\bfseries}
\titlespacing{\author}{-5pt}{-5pt}{-5pt}[-5pt]

\makeatletter
\renewcommand\subsubsection{\@startsection{subsubsection}{3}{\z@}
                                     {-3.25ex\@plus -1ex \@minus -.2ex}
                                     {-1.5ex \@plus -.2ex}
                                     {\normalfont\normalsize\bfseries}}
\renewcommand\subsection{\@startsection{subsection}{3}{\z@}
                                     {-3.25ex\@plus -1ex \@minus -.2ex}
                                     {-1.5ex \@plus -.2ex}
                                     {\normalfont\normalsize\bfseries}}                                     
\makeatother

\usepackage{soul}

\usepackage{scalerel}[2016/12/29]

\DeclareFontFamily{U}{solomos}{}
\DeclareErrorFont{U}{solomos}{m}{n}{10}
\DeclareFontShape{U}{solomos}{m}{n}{
  <-> s*[1.1]  gsolomos8r
}{}

   \usepackage{stmaryrd}

\usepackage{tikz}
\usetikzlibrary{arrows.meta}

\usepackage{indentfirst}

\let \savenumberline \numberline
\def \numberline#1{\savenumberline{#1.}}

\usepackage{xurl}
\usepackage{scrextend}

\usepackage{upgreek}

\usepackage{etoolbox}
\patchcmd{\tableofcontents}{\@starttoc}{\vspace{-0.3cm}\@starttoc}{}{}

\usepackage{nicematrix}

\newcounter{Chapcounter}

\newcommand{\chapter}[1] 
{ {\centering          
  \addtocounter{Chapcounter}{1} \Large \underline{\sffamily \texorpdfstring{\textbf{  Chapter \theChapcounter: ~#1}}{Lg}} }   
  \addcontentsline{toc}{section}{ \color{blue} \texorpdfstring{Chapter ~}{Lg}\theChapcounter.\texorpdfstring{~~}{Lg} #1 }    
}

\newcommand{\appendixbig}[1] 
{ {\centering          
   \Large \underline{\sffamily \textbf{  Appendices}} }   
  \addcontentsline{toc}{section}{ \color{blue} Appendices}    
}

\usepackage[hyperfootnotes=]{hyperref}
\hypersetup{
    colorlinks=true,
    linkcolor=blue,
    filecolor=magenta,      
    urlcolor=cyan,
    breaklinks=true
    }

\begin{document}

\title{Mechanics on flag manifolds\footnote{Prepared as a contribution to `SQS-24'.}}

\author{Andrew Kuzovchikov$^{\,a,b,}$\footnote{E-mail: andrkuzovchikov@mail.ru}
\\  \vspace{0cm}  \\
{\small $a)$ \emph{Steklov
Mathematical Institute of Russian Academy of Sciences,}} \\{\small \emph{Gubkina str. 8, 119991 Moscow, Russia} }\\
{\small $b)$ \emph{Institute for Theoretical and Mathematical Physics,}} \\{\small \emph{Lomonosov Moscow State University, 119991 Moscow, Russia}}
}

\date{}

{\let\newpage\relax\maketitle}

\maketitle

\vspace{0cm}
\textbf{Abstract.} We study the connection between $\mathrm{SU}(n)$ spin chains and one-dimensional sigma models on flag manifolds. Using this connection, we calculate the spectrum of the Laplace-Beltrami operator and geodesics for a particular class of metrics on $\mathbb{CP}^1$ and $\mathcal{F}_3$, which is a manifold of complete flags in $\mathbb{C}^3$.

\section{Introduction}

The connection between sigma models and spin chains was first explored in the seminal work of Haldane \cite{HaldaneNonlin}, which focused on a two-dimensional model on the two-sphere $\mathcal{S}^2$ and an $\mathrm{SU}(2)$ spin chain. This connection was later extended to flag manifolds and the $\mathrm{SU}(n)$ group in \cite{BykLagEmb,Bykov:2012am}. In this paper, we explore a one-dimensional version of this connection in the context of complete flag manifolds. Our aim is to use this connection to solve quantum and classical mechanics problems on these manifolds, specifically by calculating the spectrum of the Laplace-Beltrami operator and finding geodesics. 

We start with the definition of a complete flag manifold, which a homogeneous space of $\mathrm{SU}(n)$. It can be represented as 
\begin{align}
    \mathcal{F}_n := \frac{\mathrm{SU}(n)}{\mathrm{S}\left(\mathrm{U}(1)^{\times n}\right)}\,.
\end{align}
$\mathcal{F}_n$ can be parameterized by a set of mutually orthogonal vectors $\{\mathbf{u}_i \in \mathbb{CP}^{n-1}\}_{i=1}^n$. The parametrization defines a Lagrangian embedding of $\mathcal{F}_n$ into $\left(\mathbb{CP}^{n-1}\right)^{n}$, first studied in \cite{Bykov:2012am}. In present work, we will focus on two examples of complete flag manifolds: $\mathcal{F}_2 \simeq \mathbb{CP}^1$ and $\mathcal{F}_3$.
 
\section{Case of $\mathbb{CP}^1$}
	We start by reviewing some basic properties that are shared by a mechanical problem on $\mathcal{S}^2 \simeq \mathbb{CP}^1$. 
	
	Let us begin with quantum mechanics (QM). A well-known result from representation theory states that the space of square-integrable functions on $\mathcal{S}^2$ can be decomposed into a sum of $\mathrm{SU}(2)$ irreducible representations \cite{Isaev:2018xcg}:
	\begin{align}
		L^2\left(\mathcal{S}^2\right) = \lim\limits_{p \to \infty} \mathrm{T}^{\,p/2} \otimes \mathrm{T}^{\,p/2} = \lim\limits_{p \to \infty} \bigoplus\limits_{j=0}^{p} \mathrm{T}^{\,j}\,, \label{L2SphereDecomp}
	\end{align} 
	where $\mathrm{T}^{\,p/2}$ is a $\mathrm{SU}(2)$ irreducible representation of spin $p/2$. This follows from the fact that the spherical harmonics $\mathrm{Y}^{l}_{m}$ form an orthonormal basis for $L^2\left(\mathcal{S}^2\right)$. We note that the set of $\mathrm{Y}^{l}_{m}$ with fixed $l$ forms $\mathrm{T}^l$. 
	
	The standard Hamiltonian $\mathrm{H}$ of QM on $\mathcal{S}^2$ is simply $-\triangle$, where $\triangle$ is a Laplace-Beltrami operator for usual `round' metric on a sphere. For fixed $l$ and $m$, $\mathrm{Y}^{l}_{m}$ is an eigenfunction of $-\triangle$ with an eigenvalue $\Lambda_{l}:=l(l+1)$. Thus, the spectrum of $\mathrm{H}$ is $\{\Lambda_j\}_{j=0}^{\infty}$, where each $\Lambda_{l}$ has multiplicity $\mu_l = 2l+1$. 
	
	Now, we move on to classical mechanics. Let us specify the~parametrization described in the Introduction for the case of $\mathcal{S}^2$. The point on $\mathcal{S}^2 \simeq \mathbb{CP}^1$ can be parameterized by two vectors, $\{\mathbf{u}_i \in \mathbb{CP}^1\}_{i=1}^2$, subject to the additional orthogonality condition
	\begin{align}
		\Bar{\mathbf{u}}_1 \circ \mathbf{u}_2 = \sum_{j=1}^{2} \Bar{\mathbf{u}}^{j}_1 \mathbf{u}^{j}_2 =  0\,,
	\end{align}
	where $\{\mathbf{u}_i^j\}_{j=1}^2$ are components of $\mathbf{u}_i$. For simplicity, we also normalize vectors, i.e. $\Bar{\mathbf{u}}_i \,\circ\, \mathbf{u}_j = \delta_{ij}$. Using $\mathbf{u}_j$'s vectors, we can write the usual `round' metric as 
	\begin{align}
		\mathrm{d}s^2 = |\Bar{\mathbf{u}}_1 \circ \mathrm{d} \mathbf{u}_2|^2\,. \label{roundMetrS2}
	\end{align}
	We can easily calculate geodesics for the metric (\ref{roundMetrS2}). They have the form
	\begin{align}
		\mathsf{U}(t) := 
		\begin{pmatrix}
			\mathbf{u}_1(t) & \mathbf{u}_2(t) 
		\end{pmatrix}
		 = \mathsf{U}(0) \times \exp\left[ \,t
		 \begin{pmatrix}
		 	0 & a_0 \\
		 	-\Bar{a}_0 & 0
		 \end{pmatrix} 
		 \right]\,, \label{sphereGeod}
	\end{align}
	where $a_0 := \Bar{\mathbf{u}}_1 \,\circ\, \dot{\mathbf{u}}_2(0)$.
	
	One might ask: is there another system that exhibits similar behavior at the quantum and classical levels? The answer is `yes', and we will discuss it now.
	
	We start with a simple $\mathrm{SU}(2)$ spin chain consisting of only two sites. The Hamiltonian for the system is chosen as
	\begin{align}
		\mathrm{H}_{\text{spin}} := \frac{1}{2}\left(\mathrm{S}^a_1 + \mathrm{S}^a_2\right)^2 - \mathrm{const} = \mathrm{S}^a_1 \mathrm{S}^a_2\,,
	\end{align}
	where $\mathrm{S}^a_i$ ($a=1,2,3$) are generators of $\mathfrak{su}(2)$ in the representation $\mathrm{T}^{\,p/2}$ and we assume summation w.r.t. repeated indices. The Hamiltonian $\mathrm{H}_{\text{spin}}$, up to a constant, is a quadratic Casimir operator of $\mathfrak{su}(2)$, acting in the space $\mathrm{V}(p) := \mathrm{T}^{\,p/2} \otimes \mathrm{T}^{\,p/2}$. Using the Perelomov-Popov formula \cite{PerelomovPopov}, one can easily calculate the spectrum of $\mathrm{H}_{\text{spin}}$, which is given by $\{\Lambda_j\}_{j=0}^p$, with the same multiplicities as in the sphere case. Therefore, quantum mechanics on $\mathcal{S}^2$ can be recovered from the spin chain in the limit of large $p$.
	
	At the classical level, we expect that the solutions to the e.o.m. for the spin chain will somehow correspond to geodesics on $\mathbb{CP}^1$. We begin with the action for the spin chain\cite{Affleck2022}
	\begin{align}
		\mathsf{S} = \int \mathrm{d}t \left( i\,\Bar{\mathbf{z}}_1 \circ \dot{\mathbf{z}}_1 + i\,\Bar{\mathbf{z}}_2 \circ \dot{\mathbf{z}}_2 - |\Bar{\mathbf{z}}_1\circ \mathbf{z}_2|^2 \right), \label{actionSpinSU(2)}
	\end{align}
	where $\mathbf{z}_i \in \mathbb{CP}^1$ and we impose the constraint $\Bar{\mathbf{z}}_i \circ \mathbf{z}_i = p$ for $i=1,2$. The solution to the e.o.m. of the classical problem is given by the following expression
	\begin{align}
		\mathsf{Z}(t) := \begin{pmatrix}
			\mathbf{z}_1 & \mathbf{z}_2 
		\end{pmatrix}
		 = \mathsf{Z}(0) \times \exp\left[ \,t
		 \begin{pmatrix}
		 	0 & A_0 \\
		 	-\Bar{A}_0 & 0
		 \end{pmatrix} 
		 \right]\,,
	\end{align}
	where $A_0 := \Bar{\mathbf{z}}_1 \circ \mathbf{z}_2 (0)$. Due to the Cauchy-Schwartz inequality, we have $|A_0|^2 \leq p^2$
	
	We note that the solution has a similar form to the geodesic (\ref{sphereGeod}) but with a restricted momentum, which is $A_0$. This is not a coincidence. Indeed, let us take a closer look at (\ref{actionSpinSU(2)}). Suppose, $\mathsf{Z}$ is non-degenerate. Then, we can use the polar-decomposition theorem $\mathsf{Z} = \mathsf{U} \mathsf{H}$, where $\mathsf{U} = \mathsf{Z}\left(\mathsf{Z}^{\dagger}\mathsf{Z}\right)^{-1/2}$ is unitary and $\mathsf{H} = \left(\mathsf{Z}^{\dagger}\mathsf{Z}\right)^{1/2}$ is a Hermitian positive-definite matrix. Using the decomposition, the action can be rewritten as follows 
	\begin{align}
		\mathsf{S} = \int \mathrm{d}t \left[\, i\, \frac{1}{2} \frac{\mathrm{d}}{\mathrm{d}t}\mathrm{Tr} \left(\mathsf{H}^2\right) + i \mathrm{Tr} \left(\mathsf{K}\mathsf{U}^{\dagger} \dot{\mathsf{U}}\right)- \mathsf{K}_{12}\mathsf{K}_{21} \right],
	\end{align} 
	where $\mathsf{K} := \mathsf{H}^2$ and $\mathsf{K}_{ij}$'s are matrix elements of $\mathsf{K}$. The first term vanishes due to constraints. The two remaining terms form a standard first-order action for the geodesic flow on a sphere. However, $\mathsf{K}_{12}$ is not arbitrary, as $\mathsf{H}$ is positive-definite. In fact, the restrictions on $\mathsf{K}_{12}$ are lifted, when $p \to \infty$.  Thus, in the limit, we can integrate over $\mathsf{K}_{12}$ and get
	\begin{align}
		\mathsf{S} = \int \mathrm{d}t\,\, \dot{\Bar{\mathbf{u}}}_1 \circ \mathbf{u}_2 \Bar{\mathbf{u}}_2 \circ \dot{\mathbf{u}}_1 \,.
	\end{align}
	This is the action for a free particle on $\mathcal{S}^2$. In fact, we can apply the above discussion to the quantum case. Indeed, let us consider the twisted partition function
		\begin{align}
			\mathcal{Z}(\beta | g) := \mathrm{Tr}_{\mathrm{V}(p)}\left(\,g \,e^{-\beta\, \mathrm{H}_{\text{spin}}}\right)\,,
		\end{align}
		where $g \in \mathrm{SU}(2)$ is in an appropriate representation. We are interested in $\mathcal{Z}(\beta | g)$, as it contains information about the spectrum of the quantum system.  
		
        Using the path integral technique and the same arguments as in the classical case, one can find that, in the limit $p \to \infty$, $\mathcal{Z}(\beta | g)$ approaches the twisted partition function for a particle on a sphere.

        Thus, we have shown that the considered $\mathrm{SU}(2)$ spin system is a truncation of a one-dimensional system on a sphere.

	\section{Case of $\mathcal{F}_{3}$}
	
	In this section, we will briefly discuss the relationship between flag manifolds and spin chains in the case of $\mathcal{F}_3$ and $\mathrm{SU}(3)$ spin chains. 
    
    As in the sphere case there is a straightforward connection between one-dimensional sigma model on $\mathcal{F}_{3}$ and $\mathrm{SU}(3)$ spin chain containing three sites~\cite{Bykov2024}. The Hamiltonian of the appropriate spin system is
    \begin{align}
        \mathrm{H}_{\text{spin}} = \alpha\, \mathrm{S}^a_1 \mathrm{S}^a_2 + \beta\, \mathrm{S}^a_2 \mathrm{S}^a_3 + \gamma\, \mathrm{S}^a_1 \mathrm{S}^a_3 + \mathrm{const},
    \end{align}
    where $\alpha, \beta, \gamma$ are real, non-negative numbers. The generators $\mathrm{S}^a_i$ ($a=1,\ldots,8$) of the $\mathfrak{su}(3)$ algebra are taken in the $p$-th symmetric power of the associated fundamental representation, $\mathrm{Sym}(p)$. The full Hilbert space is $\mathrm{V}(p) = \mathrm{Sym}(p)^{\otimes 3}$. 

    The Hamiltonian corresponds to the following metric on $\mathcal{F}_{3}$
    \begin{align}
        \mathrm{d}s^2 = \frac{1}{\alpha} |\Bar{\mathbf{u}}_1 \circ \mathrm{d} \mathbf{u}_2|^2 + \frac{1}{\beta} |\Bar{\mathbf{u}}_1 \circ \mathrm{d} \mathbf{u}_3|^2 + \frac{1}{\gamma} |\Bar{\mathbf{u}}_2 \circ \mathrm{d} \mathbf{u}_3|^2\,,\label{SU3metric}
    \end{align}
    where $\mathbf{u}_i \in \mathbb{CP}^2$ and $\Bar{\mathbf{u}}_i\,\circ\, \mathbf{u}_j = \delta_{ij}$ for $i,j=1,2,3$.
    
    As in the sphere case, the spectrum of the Laplace-Beltrami operator for the metric (\ref{SU3metric}) can be restored as the spectrum of $\mathrm{H}_{\text{spin}}$ in the limit $p \to \infty$. It turns out that the spectrum of $\mathrm{H}_{\text{spin}}$ can be explicitly calculated in the case of $\beta = \gamma$:
    \begin{prop}[\cite{Bykov2024}]
        The eigenvalues of $\mathrm{H}_{\text{spin}}$, for a fixed value of $p$, take the form
        \begin{gather}\label{F123spectrum}
            \Lambda_k = \frac{\beta}{2}C_2(p^B_1,p^B_2,p^B_3)+
            \frac{\alpha - \beta}{2}
            \left[
                C_2\left(p_1^A, p^A_2, 0\right)-C_2(p,p,0)
            \right], 
        \end{gather}
        where $p_1^A+p_2^A=2p,\, p_2^A\leq p\leq p_1^A$, $p_1^B+p_2^B+p_3^B=3p,\, p_2^B\leq p_1^A\leq p_1^B,\, p_3^B\leq p_2^A \leq p_2^B$, $C_2(p_1,p_2,p_3)=\sum_{j=1}^3 s_j(s_j - 2j)$ and $s_i = p_i - \frac{1}{3} \sum_{j=1}^3 p_j$.
    \end{prop}
    When $\alpha = \beta = \gamma $, the metric (\ref{SU3metric}) is called `normal'. In this case, the spectrum of the Laplace-Beltrami operator on $\mathcal{F}_{3}$ was studied in \cite{Yamaguchi}. 

    We can also calculate geodesics when $\beta = \gamma$:
    \newpage
    \begin{prop}[\cite{Bykov2024}]
        In this case, the general solution to the geodesic equation on $\mathcal{F}_3$ has the form:
        \begin{align}
            &\mathsf{U}(t) := \begin{pmatrix}
                \mathbf{u}_1 & \mathbf{u}_2 & \mathbf{u}_3
            \end{pmatrix}(t)= \mathsf{U}(0) \times \exp\left[-i\beta\begin{pmatrix}
        0 & a_0 & b_0 \\
        \Bar{a}_0 & 0 & c_0 \\
        \Bar{b}_0 & \Bar{c}_0 & 0
    \end{pmatrix} t \right]\times \mathsf{G}(t)\,,\\
    &\text{where} \quad 
    \mathsf{G}(t) :=
    \begin{pmatrix}
        \exp 
        \left[
        i\,\left(\beta - \alpha\right)
        \begin{pmatrix}
            0 & a_0 \\
            \Bar{a}_0 & 0
        \end{pmatrix}t
        \right] & \\
        & 1
    \end{pmatrix}\,
        \end{align}
        and $a_0,b_0,c_0$ are the initial data.
    \end{prop}
    The last proposition coincides with the result of the work \cite{Souris}. In fact, the described connection between spin chains and flag manifolds can be generalized to the supersymmetric case \cite{BKK}.

\label{sec:acknowledgement}
\section{Acknowledgement}
This work is supported by the Russian Science Foundation grant № 22-72-10122 (\href{https://rscf.ru/en/project/22-72-10122/}{\emph{https://rscf.ru/en/project/22-72-10122/}}). I would like to thank Viacheslav Krivorol, Mikhail Markov, Anton Selemenchuk, Sergey Derkachov and especially Dmitri Bykov for useful discussions, comments on the work, and inspiration.

\vspace{1cm}    
    \setstretch{0.8}
    \setlength\bibitemsep{5pt}
    \printbibliography

@article{Bykov:2012am,
    author = "Bykov, Dmitri",
    title = "{The geometry of antiferromagnetic spin chains}",
    eprint = "1206.2777",
    archivePrefix = "arXiv",
    primaryClass = "hep-th",
    reportNumber = "NORDITA-2012-47",
    doi = "10.1007/s00220-013-1702-5",
    journal = "Commun. Math. Phys.",
    volume = "322",
    pages = "807-834",
    year = "2013",
    SLACcitation   = "%%CITATION = ARXIV:1206.2777;%%"
}

@article{HaldaneNonlin,
    author = "Haldane, F. D. M.",
    title = "{Nonlinear field theory of large spin Heisenberg antiferromagnets. Semiclassically quantized solitons of the one-dimensional easy Axis Neel state}",
    doi = "10.1103/PhysRevLett.50.1153",
    journal = "Phys. Rev. Lett.",
    volume = "50",
    pages = "1153--1156",
    year = "1983"
}

@article{BykLagEmb,
    author = "Bykov, Dmitri",
    title = "Haldane limits via Lagrangian embeddings",
    eprint = "1104.1419",
    archivePrefix = "arXiv",
    primaryClass = "hep-th",
    journal = "Nuclear Physics B",
    volume = "855",
    pages = "pp. 100 - 127",
    doi = "10.1016/j.nuclphysb.2011.10.005",
    year = "2012"
}

@article{Souris,
author = {Souris, Nikolaos},
title = {Geodesics as products of one‐parameter subgroups in compact lie groups and homogeneous spaces},
year = {2023},
pages = {2609-2625},
volume = {296},
number = "6",
journal = {Mathematische Nachrichten},
doi = {10.1002/mana.202000282}
}

@article{Bykov2024,
doi = {10.1088/1361-6382/ad7189},
year = {2024},
publisher = {IOP Publishing},
volume = {41},
number = {20},
pages = {205009},
author = {Dmitri Bykov and Andrew Kuzovchikov},
title = {The classical and quantum particle on a flag manifold},
journal = {Class. Quant. Grav.},
eprint = "2404.15900",
archivePrefix = "arXiv",
primaryClass = "hep-th",
SLACcitation   = "%%CITATION = ARXIV:2404.15900;%%"
}

@Article{Yamaguchi,
 Author = {Yamaguchi, Satoru},
 Title = {Spectra of flag manifolds},
 FJournal = {Memoirs of the Faculty of Science. Series A. Mathematics},
 Journal = {Mem. Fac. Sci., Kyushu Univ., Ser. A},
 ISSN = {0373-6385},
 Volume = {33},
 Pages = {95--112},
 Year = {1979},
 Language = {English},
 DOI = {10.2206/kyushumfs.33.95},
 zbMATH = {3629763},
 Zbl = {0405.53025}
}

@article{PerelomovPopov,
author={A.~M.~Perelomov and V.~S.~Popov}, 
title = "Casimir operators for semisimple Lie groups",
journal = "Math. USSR-Izv.",
year = "1968",
volume="2",
number="6",
pages="1313--1335",
doi="10.1070/IM1968v002n06ABEH000731"
}

@article{Affleck2022,
   author="Affleck, Ian and Bykov, Dmitri and Wamer, Kyle",
   title="Flag manifold sigma models",
   eprint="2101.11638",
   archivePrefix  = "arXiv",
   primaryClass   = "hep-th",
   volume="953",
   doi="10.1016/j.physrep.2021.09.004",
   journal="Physics Reports",
   publisher="Elsevier BV",
   pages="1–93",
   year="2022",
   SLACcitation   = "%%CITATION = ARXIV:2101.11638;%%"
}

@book{Isaev:2018xcg,
    author = "Isaev, AlexeyP. and Rubakov, ValeryA.",
    title = "Theory of Groups and Symmetries",
    doi = "10.1142/10898",
    publisher = "WSP",
    year = "2018"
}

@article{BKK,
    author = "Bykov, Dmitri and Krivorol, Viacheslav and Kuzovchikov, Andrew",
    title = "{Oscillator Calculus on Coadjoint Orbits and Index Theorems}",
    eprint = "2412.21024",
    archivePrefix = "arXiv",
    primaryClass = "hep-th",
    year = "2024",
    doi = {10.48550/arXiv.2412.21024}
}

\end{document}